\documentclass[11pt]{article}
\usepackage[pdftex]{graphicx}
\usepackage{hyperref}
\usepackage{cite}
\usepackage{amsmath,amsfonts,amssymb,amsthm,nccmath,latexsym,mathtools}
\usepackage{xcolor}
\usepackage{amsmath}
\usepackage{cancel} %to strike out obliquely
\usepackage{ amssymb }
\usepackage{footnote}

\def\s{\sigma}

\def\no{\nonumber}
\def\a{\alpha}
\def\b{\beta}
\def\g{\gamma}
\def\d{\delta}
\def\s{\sigma}

\def\p{\partial}

\def\t{\tilde}

\def\l{\lambda}
\def\be{\begin{equation}}
\def\ee{\end{equation}}
\def\ba{\begin{align}}
\def\ea{\end{align}}

\def\z{\zeta}
\def\etc{{\it etc.}}
\def\X{\mathcal{X}}
\def\Y{\mathcal{Y}}
\def\ie{{\it i.e.~}}
\begin{document}
\title{{\bf{\Large A general framework to study the extremal phase transition of black holes}}}
\author{
{\bf{\normalsize Krishnakanta Bhattacharya$^{a}$}}\footnote {\color{blue} krishnakanta@iitg.ac.in}, \
{\bf{\normalsize Sumit Dey$^{a}$}}\footnote {\color{blue} dey18@iitg.ac.in}, \
{\bf{\normalsize Bibhas Ranjan Majhi$^{a}$}}\footnote {\color{blue} bibhas.majhi@iitg.ac.in} \ and \ {\bf{\normalsize Saurav Samanta$^{b}$}}\footnote {\color{blue} srvsmnt@gmail.com}\\
$^a$Department of Physics, Indian Institute of Technology Guwahati, Guwahati 781039, Assam, India\\
$^b$Department of Physics, Bajkul Milani Mahavidyalaya, P. O. - Kismat Bajkul,\\
 Dist. - Purba Medinipur, Pin - 721655, India
}
%\date{\today}
\maketitle
%%%%%%%%%%%%%%%%%%%%%%%%%%%%%%%%%%%%%%%%%%%%%%%%
\begin{abstract}
We investigate the universality of some features for the extremal phase transition of black holes and unify all the approaches which have been applied in different spacetimes. Unlike the other existing approaches where the information of the spacetime and its dimension is directly used to get various results, we provide a general formulation in which those results are obtained for any arbitrary black hole spacetime having an extremal limit. Calculating the second order moments of fluctuations of some thermodynamic quantities we show that, the phase transition occurs only in the microcanonical ensemble. Without considering any specific black hole we calculate the values of critical exponents for this type of phase transition. These are shown to be in agreement with the values obtained earlier for metric specified cases. Finally we extend our analysis to the geometrothermodynamics (henceforth GTD) formulation. We show that for any black hole, if there is an extremal point, the Ricci scalar for the Ruppeiner metric must diverge at that point.

\end{abstract}

%%%%%%%%%%%%%%%%%%%%%%%%%%%%%%%%%%%%%%%%

%%%%%%%%%%%%%%%%%%%%%%%%%%%%%%%%%%%%%%%%%%%%%%%%%%%%%%%%%%%%%%%%%%%%%%%%%%%%%%%
\section{Introduction}
The remarkable discovery of Bekenstein \cite{Bekenstein:1973ur} and Hawking \cite{Hawking:1974sw} in the seventies laid the foundation of black hole thermodynamics, which has been the subject of ardent research in the following decades till date. Identifying the thermodynamic parameters (such as entropy, temperature, energy \etc ) from the geometrical quantities of the black hole sapcetime (such as the area of the horizon, surface gravity of the black hole horizon \etc), four laws of black hole mechanics were formulated in 1973 \cite{Bardeen:1973gs}. These works clearly imply the existence of thermodynamic structure of the black hole horizon. Since then, many thermodynamic phenomena have been observed in black hole spacetime. The study of phase transition, which is an important phenomenon in ordinary thermodynamics, has also been explored in black hole mechanics since seventies. It was introduced by Davies \cite{Davies:1978mf} and subsequently followed by many other researchers \cite{Lousto:1993yr, Lousto:1994cz, Lousto:1994jd, Muniain:1995ih}. Davies endorsed that a black hole goes through a second order phase transition when it passes through a point (Davies' point) where the heat capacity becomes infinitely discontinuous. However, later Kaburaki {\it et al.} \cite{Kaburaki:1993ah, Katz:1993up, Kaburakipla, Kaburakigrg} claimed that Davies' point is not a critical point. Instead, it is merely a turning point, where stability changes.

Although, Davies' claim was later falsified, other groups argued that a second order phase transition indeed takes place when a non-extremal black hole transforms to an extremal one and the extremal limit was identified as a critical point. It was first concluded by Curir in \cite{curir1, curir2}. Later Pav$\acute{\textrm{o}}$n and Rub$\acute{\textrm{i}}$ \cite{Pavon:1988in, Pavon:1991kh} calculated second order moments of fluctuation of mass, angular momentum \etc~using Landau-Lifshitz hydrodynamic fluctuation theory (see chapter 17 of \cite{Landaufluid})  and have shown that those second order moments diverge in the extremal limit of Kerr and Reissner-Nordstr$\ddot{\textrm{o}}$m (RN) black holes but those moments are finite in the non-extremal limit and for the Schwarzschild black hole. Also, those second order moments remain finite at the Davies' point. Both the analysis are in agreement with each other and suggest that the extremal limit of the black hole is a critical point, and the divergence of second order moments of fluctuation should signal a second order phase transition of the black holes which are changing from its non-extremal phase to the extremal phase. Later, this phase transition in the extremal limit has been rigorously studied for different (Kerr-Newman \cite{Kaburakigrg}, BTZ \cite{Cai:1996df, Cai:1998ep, Wei:2009zzf} \etc) black holes and critical exponents were obtained. These exponents satisfy the well known scaling laws \cite{Stanley1, book} of thermodynamics. 

 The works, which are mentioned above, are performed in different spacetimes to come to the same central conclusion that the extremal limit is a critical point and the transformation from a non-extremal to an extremal black hole is a second order phase transition. Moreover, in those cases, the information of the spacetime has directly been used to obtain the results.  One question naturally appears: is it really necessary to start with a particular spacetime to reach this conclusion? The results present in different papers suggest us to believe that probably the conclusion is true irrespective of spacetime metric and its dimension. But till now there has not been any such proof. Moreover, there are few major questions which has not been addressed properly. Some of these are: Are the critical exponents universal? Is the effective spatial dimension one in every extremal black holes \etc  \ In this paper we address all these issues systematically.
 
 Our analysis is valid for all the black holes which are extremal at certain limit. Without introducing any particular spacetime we show that the transformation of black hole from non-extremal to extremal is a second order phase transition with the extremal limit being the critical point. To prove that, we calculate the second order moments of fluctuation modes of some thermodynamic quantities using equilibrium fluctuation theory of statistical mechanics \cite{Landau, Kaburakipla, Kaburakigrg} and show that those moments diverge in the microcanonical ensemble. Thereby we show that the phase transition is well described only by the micro-canonical ensemble instead of the canonical or the grand canonical ensembles. Later, we proceed our analysis to obtain the values of critical exponents in a general way. These exponents match with the results, obtained earlier by considering the explicit form of the spacetime. Also these have been shown to satisfy the scaling laws. We emphasize that in our whole analysis the only underlying information one requires is: { \it one should consider the particular class of black hole spacetimes which exhibit such non-extremal to extremal transition at certain limit} and additionally, {\it the thermodynamics of those black holes are governed by the usual first law of black hole mechanics at the non-extremal limit}. 
 
 We also analyze another interesting aspect in our paper. It is known for a long time that classical thermodynamics can also be studied by geometric method. This is the GTD formulation. In Weinhold's approach the metric is defined as the Hessian of the internal energy and in the Ruppeiner's approach the metric is defined as the Hessian of the entropy. It has been shown that Ruppeiner curvature scalar diverges at the extremal limit of the BTZ black hole\cite{Cai:1998ep, Wei:2009zzf}. In the present paper we have proved this result for any arbitrary black hole which has an extremal point.
 
 Very recently it has been claimed that neither the Weinhold nor Ruppeiner formulation is Legendre invariant and, hence, they are inappropriate to analyze the thermodynamics. So, we proceed one step further to find the thermodynamic behaviour at the extremal point using Legendre-invariant metric. We do this for two Quevedo GTD metrics and find that the Ricci scalar for both of those metrics are finite at the extremal point. Thus, our work connects all the previous diverse conclusions about extremal phase transitions, all of which are black hole specific. In this sense, our work is unique and fills an important gap in the literature.   
 
 Before we proceed further, let us mention the organization of our paper. In the second section we discuss the black hole thermodynamics at the extremal point without using any particular form of spacetime. Second order moments of fluctuation are calculated for microcanonical, canonical and grand canonical ensembles in three subsections. It is observed that the phase transition is compatible with the first ensemble. Next section is dedicated to calculate the values of different critical exponents. Then in section \ref{sec4}, thermogeometric analysis has been performed separately for Weinhold, Ruppeiner and two Legendre invariant metrics. It is shown that the curvature scalar diverges only for the Ruppeiner metric. Finally, in the last section, we draw conclusions of our work.

\section{Thermodynamic analysis of extremal point in different ensembles}
We have already mentioned that, the extremal phase transition is regarded as a second order phase transition. This was first claimed by Curir \cite{curir1, curir2}. According to  Pav$\acute{\textrm{o}}$n and Rub$\acute{\textrm{i}}$ \cite{Pavon:1988in, Pavon:1991kh}, the divergence of the second order moments of fluctuations of thermodynamic quantities is a signature of this phase transition. Following this argument, here we calculate these second order moments in different ensembles. We show that, only in microcanonical ensemble extremal limit of black hole (if it exists) is a second order phase transition.

Here, we calculate the second order moments using the well-defined equilibrium fluctuation theory of statistical mechanics. In that case, the required thermodynamical quantities are obtained from Massieu function, which are the Legendre transformations of the entropy. In that formalism, the state of a given environment is completely characterized by the Massieu function \cite{Kaburakipla, Kaburakigrg} $\Phi$, whose variation is given by
 \begin{align}
 d\Phi=\X_id\Y^i~. \label{MASSIEU}
 \end{align}
 Here, the summation convention has been adopted. In the above relation, Massieu function is a function of the intrinsic variables $\Y^i$. $\X_i$ which is the conjugate variables of $\Y^i$, is defined as $\X_i=(\p \Phi/\p \Y^i)_{\bar{\Y}^i}$~. In our notation, $\bar{\Y}^i$ is the set of all intrinsic variables excluding $\Y^i$~. {\it Throughout our analysis, a bar overhead will imply similar thing}. Now for a given environment, the spontaneous fluctuation from the equilibrium occurs only in the conjugate variables $\X_i$. This is because, the reservoirs are considered to be large compared to the system and as a result, the intrinsic variables are fixed. Then the probability of the deviation from the equilibrium is proportional to $\textrm{exp}[-\Sigma\lambda_i(\d \X^i)^2/(2k_B)]$ \cite{Kaburakigrg}, where $k_B$ is the Boltzmann constant. The eigenvalues of the fluctuation modes are defined as
\begin{align}
\lambda_i=\frac{\p \Y_i}{\p \X^i}\Big|_{\bar{\Y}^i}=\Big(\frac{\p^2\Phi}{\p \Y^{i^{2}}}\Big)^{-1}_{\bar{\Y}^i}~.
\end{align}
 Here it should be mentioned that the probability is accurate only up to the second order. The averages of modes of fluctuations always vanish \cite{Landau} and the second order moments are given by
 \begin{align}
 \mathcal{M}_{ij}=\langle\d\X_i\d\X_j\rangle=k_B\Big(\frac{\p^2\Phi}{\p \Y^{i2}}\Big)_{\bar{\Y}^i}\d_{ij}=\frac{k_B}{\lambda_i}\d_{ij}~. \label{SECMOM}
 \end{align}
 In the following analysis, we investigate the behaviour of these quantities in each ensemble. Since the extremal limit is not a turning point \cite{Kaburakigrg}, the divergence of the second order moments will imply the presence of second order phase transition.
 
 %---------------------------------------------------------------------------------------------------
 
\subsection{Microcanonical ensemble}
Let us consider an isolated black hole by definition which exchanges nothing with the environment. In this case, the proper Massieu function $\Phi_1$ is the entropy $S$. Its change is given by the first law of black hole mechanics{\footnote{This is one of the inputs of our present discussion; whereas the other input is the existence of extremal limit in the black hole thermodynamics.}}: 
\begin{align}
dS=\b dM-\t X^idY_i~, \label{ENTROPY}
\end{align}
where $\b=1/T$ and $\t X^i=\b X^i$~. According to our notations $X^i$ are potential, angular velocity \etc, whereas $Y_i$ are charge, angular momentum \etc~Then the eigenvalues of the fluctuations are given by
\begin{align}
\lambda^{(1)}_M=\Big(\frac{\p^2 S}{\p M^2}\Big)_{Y_i}^{-1}=\Big(\frac{\p M}{\p \b}\Big)_{Y_i}=-T^2C_Y~ \label{LAMBM}
\end{align}
and 
\begin{align}
\lambda^{(1)}_{Y_i}=\Big(\frac{\p^2 S}{\p Y_i^2}\Big)_{M, \bar Y_i}^{-1}=-\Big(\frac{\p Y_i}{\p \t X^i}\Big)_{M, \bar Y_i}=-TI_M^{(i)}~. \label{LAMBYI}
\end{align}
Here we used the following definitions: $C_Y=(\p M/\p T)_{Y_i}=-\b^2(\p M/\p \b)_{Y_i}$ and $I_M^{(i)}=(\p Y_i/\p X^i)_{M, \bar Y_i}=\b(\p Y_i/\p \t X^i)_{M, \bar Y_i}$~. Therefore the second order moments are given by
\begin{align}
\langle\d\b\d\b\rangle=k_B\Big(\frac{\p^2 S}{\p M^2}\Big)_{Y_i}=-k_B \frac{\b^2}{C_Y}~ \label{DELBDELB}
\end{align}
and
\begin{align}
\langle\d \t X^i\d \t X^i\rangle=k_B\Big(\frac{\p^2 S}{\p Y_i^2}\Big)_{{M, \bar Y_i}}=-k_B \frac{\b}{I^{(i)}_M}~. \label{DELXDELX1}
\end{align}
In the following section, where we obtain the critical exponents in a general way, we show that both $(\p^2 S/\p M^2)_{Y_i}$ and $(\p^2 S/\p Y_i^2)_{{M, \bar Y_i}}$ diverge at the extremal limit (see \eqref{28} and \eqref{33}). Therefore, we can conclude from \eqref{LAMBM} and \eqref{LAMBYI} that all the eigenvalues $\lambda_M^{(1)}$ and $\lambda_{Y_i}^{(1)}$ vanish. As a result, from \eqref{DELBDELB} and \eqref{DELXDELX1} we see that all the second order moments diverge, which is the signature of phase transition. Thus, in the microcanonical ensemble, an extremal phase transition is a second order phase transition with the extremal limit being the critical point. 

%------------------------------------------------------------------------------------------------------
%%%%%%%%%%%%%%%%%%%%%%%%%%%%%%%%%%%%%%%%%%%%%%%%%%%%%%%%%%%%%%%%%%%%%%%%%%%%%%%%%

\subsection{Canonical ensemble}
In canonical ensemble, black hole can exchange only energy with the environment. The proper Massieu function ($\Phi_2$) in this ensemble is $\Phi_2=S-\b M=-\b F$, where $F=M-TS$ is the Helmholtz free energy. Note that $dF=-SdT+X^idY_i$ and $d\Phi_2=-Md\b-\t X^idY_i$. Therefore, in this case, the intrinsic variables are $\b$ and $Y_i$ whereas the conjugate quantities are $(-M)$ and $(-\t X^i)$. The eigenvalues are given by 
\begin{align}
\lambda_{\b}^{(2)}=\Big(\frac{\p^2 \Phi_2}{\p \b^2}\Big)_{Y_i}^{-1}=-\Big(\frac{\p \b}{\p M}\Big)_{Y_i}=\frac{\b^2}{C_Y}~ \label{LAMBB}
\end{align}
and
\begin{align}
\lambda^{(2)}_{Y_i}=\Big(\frac{\p^2 \Phi_2}{\p Y_i^2}\Big)_{\b, \bar Y_i}^{-1}=-\Big(\frac{\p Y_i}{\p \t X^i}\Big)_{\b, \bar Y_i}=-TI_{\b}^{(i)}~. \label{LAMBYI2}
\end{align}
In the above, we have used $I_{\b}^{(i)}=(\p Y_i/\p X^i)_{{\b}, \bar Y_i}=\b(\p Y_i/\p \t X^i)_{{\b}, \bar Y_i}$. The second order moments, in this case, are found to be 
\begin{align}
\langle\d M\d M\rangle=k_B\Big(\frac{\p^2 \Phi_2}{\p \b^2}\Big)_{Y_i}=k_B T^2C_Y~ \label{DELMDELM}
\end{align}
and 
\begin{align}
\langle\d \t X^i\d \t X^i\rangle=k_B\Big(\frac{\p^2 \Phi_2}{\p Y_i^2}\Big)_{\b, \bar Y_i}=-k_B\frac{\b}{I_{\b}^{(i)}}~. \label{DELXDELX}
\end{align}
In the appendix \ref{APPENCANO}, we show that $(\p^2 \Phi_2/\p \b^2)_{Y_i}$ vanishes and $(\p^2 \Phi_2/\p Y_i^2)_{\b, \bar Y_i}$ diverges. As a result $\lambda_{\b}^{(2)}$ in \eqref{LAMBB} diverges and $\lambda^{(2)}_{Y_i}$ in \eqref{LAMBYI2} vanishes. Also, the nature of the second order moments are evident:  $\langle\d M\d M\rangle$ of \eqref{DELMDELM} vanishes and $\langle\d \t X^i\d \t X^i\rangle$ of \eqref{DELXDELX} diverges. Therefore the extremal limit is not a critical point in the canonical ensemble.

%----------------------------------------------------------------------------------------------------

\subsection{Grand canonical ensemble} 
Finally we consider the black hole in grand canonical ensemble. It means, the black hole not only exchanges energy with the environment but also performs work on the surroundings. The proper Massieu function in this case is $\Phi_3=\Phi_2+\t X^iY_i=S-\b M+\t X^iY_i=-\b G$. Wher, $G=M-TS-X^iY_i$ is Gibbs free energy. The variation of $G$ is $dG=-SdT-Y_idX^i$ and the variation of Massieu function $\Phi_3$ is $d\Phi_3=-Md\b+Y_id\t X^i$~. Therefore in this ensemble, the intrinsic variables are $\b$ and $\t X^i$' whereas the conjugate variables are $(-M)$ and $Y_i$. The eigenvalues of the fluctuation modes are  
\begin{align}
\lambda_{\b}^{(3)}=\Big(\frac{\p^2 \Phi_3}{\p \b^2}\Big)_{\t X_i}^{-1}=-\Big(\frac{\p \b}{\p M}\Big)_{\t X_i}=\frac{\b^2}{C_{\t X}}~ \label{LAMBB2}
\end{align}
and 
\begin{align}
\lambda^{(3)}_{\t X^i}=\Big(\frac{\p^2 \Phi_3}{\p \t {X^i}^2}\Big)_{\b, \bar{\t X}^i}^{-1}=\Big(\frac{\p \t X^i}{\p Y_i}\Big)_{\b, \bar{\t X}^i}=\frac{\b}{I_{\b}^{(i)}} ~. \label{LAMBXI}
\end{align}
In the above, we have used $C_{\t X}=(\p M/\p T)_{\t X^i}=-\b^2(\p M/\p \b)_{\t X^i}$~. The second order moments in grand canonical ensemble are 
\begin{align}
\langle\d M\d M\rangle=k_B\Big(\frac{\p^2 \Phi_3}{\p \b^2}\Big)_{\t X^i}=k_B T^2C_{\t X}~ \label{DELMDELM2}
\end{align}
and 
\begin{align}
\langle\d Y_i\d Y_i\rangle=k_B\Big(\frac{\p^2 \Phi_3}{\p \t {X^i}^2}\Big)_{\b, \bar{\t X}^i}=k_B TI_{\b}^{(i)}~.  \label{DELYDELY}
\end{align}
In the appendix \ref{APPENGRAND}, we show that both $(\p^2 \Phi_3/\p \b^2)_{\t X^i}$ and $(\p^2 \Phi_3/\p \t {X^i}^2)_{\b, \bar{\t X}^i}$ vanish. As a result, we conclude that both the eigenvalues of the fluctuation modes $\lambda_{\b}^{(3)}$ and $\lambda^{(3)}_{\t X^i}$ diverge. Naturally both the second order moments $\langle\d M\d M\rangle$ and $\langle\d Y_i\d Y_i\rangle$ vanish. As a result, the extremal limit is not a second order phase transition in the grand canonical ensemble.

%-------------------------------------------------------------------------------

%%%%%%%%%%%%%%%%%%%%%%%%%%%%%%%%%%%%%%%%%%%%%%%%%%%%%%%%%%%%%%%%%%%%%%%%%
\section{Obtaining the critical exponents in a general way}
In the earlier section, we have generally shown that the extremal phase transition is indeed a second order thermodynamic phase transition in the microcanonical ensemble. In this section we obtain the values of the critical exponents in a general manner. There are several works which studied extremal criticality and obtained the critical exponents case by case. For example, in \cite{Kaburakigrg} the extremal phase transition of Kerr-Newman black hole was studied and critical exponents were obtained. Similar studies were done for BTZ black hole in \cite{Cai:1996df, Cai:1998ep, Wei:2009zzf}. In our general framework, we obtain the values of critical exponents in a metric independent way.

The critical exponts are defined for the response coefficients and for the order parameters to show how those quantities diverge near the critical point \cite{Kaburakipla2}. The response coefficients are defined as the inverse of the eigen values $\l_i$'s \cite{Kaburakipla}. For the extremal phase transition and in the microcanonical ensemble, the response coefficients are defined as:
\begin{eqnarray}
&&\z_{Y}=\Big(\frac{\p^2S}{\p M^2}\Big)\Big|_{Y_i}, \label{JAIY}
\\
&&\z_{M}^{i}=\Big(\frac{\p^2S}{\p Y_i^2}\Big)\Big|_{M, \bar Y_i} \label{JAIM}
\end{eqnarray}
In the first definition, $Y_i$ includes all the charges present in the theory, whereas, in the second definition, $\bar Y_i$ includes all the charges except $Y_i$. In classical thermodynamics, the order parameters are the difference of some extensive quantities of the two different phases. For black hole, the order parameters are defined as the difference of the conjugate quantities on the inner and the outer horizon \cite{Kaburakipla2, Su:1994zz, Cai:1996df, Cai:1997cv}. For the presence of multiple charge and angular momentum, we define the order parameters in a general manner,
\begin{align}
\eta_{Y_i}=\t X^i_{+}-\t X_{-}^i~ \label{ETA}
\end{align}
where, $\t X^i=(X^i/T)=-(\p S/\p Y_i)_{M, \bar Y_i}$ as we have defined earlier. The subscripts ``$+$'' and ``$-$'' stands for the outer horizon ($r_{+}$) and inner horizon ($r_{-}$) respectively. Now, the critical exponents are defined as \cite{Kaburakipla2}
\begin{eqnarray}
&& \z_{Y}\sim m^{-\a} \ \ \ \ \ \textrm{(for $Y_i=Y_{ic}$)} \label{ALP}
\\
&& \z_{Y}\sim y_i^{-\phi_i} \ \ \ \ \ \textrm{(for $M=M_c$ and $\bar Y_i=\bar Y_{ic}$)}\label{PHI}
\\
&& \z_{M}^{i}\sim m^{-\g_i} \ \ \ \ \textrm{(for $Y_i=Y_{ic}$)}\label{GAM}
\\
&& \z_{M}^{i}\sim y_i^{-\s_i} \ \ \ \ \ \textrm{(for $M=M_c$ and $\bar Y_i=\bar Y_{ic}$)} \label{SIG}
\\
&& \eta_{Y_i}\sim m^{\b_i} \ \ \ \ \ \ \textrm{(for $Y_i=Y_{ic}$)} \label{BET}
\\
&& \eta_{Y_i}\sim y_i^{\d_i^{-1}} \ \ \ \ \ \textrm{(for $M=M_c$ and $\bar Y_i=\bar Y_{ic}$)} \label{DEL}
\end{eqnarray}
Here we use the notation $m=1-M/M_c$ and $y_i=1-Y_i/Y_{ic}$, whereas $c$, in the subscript, signifies the corresponding values at the critical point. Remember that the critical point, in our present discussion, is the extremal point where temperature $T$ vanishes.
 
 Now we expand the mass as a function of entropy $S$ and charge $Y_i$ near the critical point. Then 
 \begin{eqnarray}
\no
&& M=a_{00}+a_{20}s^2+a_{30}s^3+a_{40}s^4+...
\\
\no 
&& \ \ \ \ \ \ +a_{01}^{(1)}y_1+a_{02}^{(1)}y_1^2+a_{03}^{(1)}y_1^3+a_{04}^{(1)}y_1^4+...\\
\no 
&& \ \ \ \ \ \ +a_{01}^{(2)}y_2+a_{02}^{(2)}y_2^2+a_{03}^{(2)}y_2^3+a_{04}^{(2)}y_2^4+...\\
&& \ \ \ \ \ \ \ \ +...+a_{11}^{(1)}s y_1 +a_{11}^{(2)}s y_2. . .  + a_{ij}^{(k)}s^iy_k^j...\label{MEXP} 
 \end{eqnarray}
Note that here $a_{10}\sim(\p M/\p S)_c=T_c=0$. Therefore, it has not appeared in the expansion of the mass. Now the contribution up to first order is
\begin{align}
\Big(\frac{\p M}{\p s}\Big)_{Y_i}\sim A_{10}s+A_{01}^{(k)}y_k ~. \label{kl}
\end{align}
Here we have rescaled the coefficients as $A_{ij}^{(k)}=(i+1)a_{i+1~j}^{(k)}$~. One can keep higher order terms in the above equation without any change of conclusion. Thus first order contribution serves our purpose. Now, we calculate $(\p^2S/\p M^2)_{Y_i}$ in the following way.
\begin{align}
\Big(\frac{\p^2S}{\p M^2}\Big)_{Y_i}\sim \Big(\frac{\p}{\p M}\Big(\frac{\p M}{\p S}\Big)^{-1}_{Y_i}\Big)_{Y_i}\sim \Big(\frac{\p}{\p M}\Big[\frac{1}{A_{10}s+A_{01}^{(k)}y_k}\Big]\Big)_{Y_i}~.
\end{align}
Therefore using \eqref{kl} we finally obtain,
\begin{align}
\Big(\frac{\p^2S}{\p M^2}\Big)\Big|_{Y_i}\sim \frac{1}{(A_{10}s+A_{01}^{(k)}y_k)^2}\frac{\p s}{\p M}\sim \frac{1}{(A_{10}s+A_{01}^{(k)}y_k)^3}~. \label{P2SPM2}
\end{align}
When $Y_i=Y_{ic}$ we find $s\sim m^{1/2}$ (from \eqref{MEXP})~. Thus from \eqref{P2SPM2}, taking the leading order contribution we get,
\begin{align}
\Big(\frac{\p^2S}{\p M^2}\Big)\Big|_{Y_i}\sim m^{-\frac{3}{2}} \ \ \ \ \ \textrm{(for $Y_i=Y_{ic}$)}\label{28}
\end{align}
Therefore from the definition of the critical exponent $\a$ (see \eqref{ALP}), we find $\a=3/2$~. 

Again when $M=M_c$ and $\bar Y_i=\bar Y_{ic}$, we obtain $s\sim y_i^{1/2}$ (from \eqref{MEXP}). Thus, from \eqref{P2SPM2} we get $(\p^2S/\p M^2)_{Y_i}\sim (A_{10}y_i^{1/2}+A_{01}^{(i)}y_i)^{-3}$. This implies that the quantity diverges as
\begin{align}
\Big(\frac{\p^2S}{\p M^2}\Big)\Big|_{Y_i}\sim y_i^{-\frac{3}{2}} \ \ \ \ \ \textrm{(for $M=M_c$ and $\bar Y_i=\bar Y_{ic}$)}.
\end{align}
Therefore from the definition \eqref{PHI}, we get $\phi_i=3/2$~.

Next we expand $Y_i$ as a function of $S$, $M$ and other charge $\bar Y_i$: 
\begin{eqnarray}
\no
&& Y_i=a_{000}+a_{200}s^2+a_{300}s^3+a_{400}s^4+... 
\\
\no 
&& \ \ \ \ \ \ +a_{010} m+a_{020} m^2+a_{030} m3+... 
\\ 
&& \ \ \ \ \ \ +...+ a_{jkl}^{(p)}s^jm^k y_{p}^l+...\label{YEXP}
\end{eqnarray}
Similar to the earlier case, here $a_{100}\sim T_c=0$. Note that $Y_p$ includes all the charges except $Y_i$. Therefore, from \eqref{YEXP} we obtain up to the first order 
\begin{align}
\frac{\p Y_i}{\p s}\Big|_{M, \bar Y_i}\sim A_{100} s+A_{010}m+A^{(p)}_{001} y_{p}~. \label{PYPS}
\end{align}
Again, we have rescaled the coefficients as $A_{jkl}^{(p)}=(j+1)a_{j+1~kl}^{(p)}$~. It should be mentioned that first order contribution is enough to serve our purpose. Now, following the similar approach as was done earlier, we obtain
\begin{align}
\frac{\p^2S}{\p Y_i^2}\Big|_{M, \bar Y_i}\sim\frac{1}{\Big(\frac{\p Y^i}{\p s}\Big)^3}\Big|_{M, \bar Y_i}\sim\frac{1}{(A_{100} s+A_{010}m+A^{(p)}_{001} y_{p})^3}. \label{P2SPY2}
\end{align}
Now, for all $Y_i=Y_{ic}$, we obtain from \eqref{YEXP} $s\sim m^{1/2}$. This when substituted in \eqref{P2SPY2} gives $(\p^2S/\p Y_i^2)_{M, \bar Y_i}\sim (A_{100} m^{1/2}+A_{010}m)^{-3}$~. Therefore, the leading order contribution gives
\begin{align}
\frac{\p^2S}{\p Y_i^2}\Big|_{M, \bar Y_i}\sim m^{-\frac{3}{2}} \ \ \ \ \ \textrm{(for $Y_i=Y_{ic}$)}~.\label{33}
\end{align}
Therefore from the definition of $\g_i$ (see \eqref{GAM}), we find $\g_i=3/2$~.

Again when $M=M_c$ and $\bar Y_i=\bar Y_{ic}$, we obtain from \eqref{YEXP} $s\sim y_i^{1/2}$. Therefore from \eqref{P2SPY2} we get the result
\begin{align}
\frac{\p^2S}{\p Y_i^2}\Big|_{M, \bar Y_i}\sim y_i^{-\frac{3}{2}} \ \ \ \ \ \textrm{(for $M=M_c$ and $\bar Y_i=\bar Y_{ic}$)}~.
\end{align}
Therefore, from the definition of the critical exponent $\s_i$ (in eq. \eqref{SIG}) we obtain $\s_i=3/2$~.

Again from \eqref{PYPS}, the leading order contribution provides 
\begin{align}
\t X^i\sim \frac{\p Y_i}{\p S}\Big|_{M, \bar Y_i}^{-1}\sim \frac{1}{A_{100}} m^{-\frac{1}{2}} \ \ \ \ \ \textrm{(for $Y_i=Y_{ic}$)}~.
\end{align}
The above equation implies 
\begin{align}
\eta_{Y_i}=\t X^i_{+}-\t X^i_{-}\sim \Big(\frac{1}{A_{100}}\Big|_{+}-\frac{1}{A_{100}}\Big|_{-}\Big) m^{-\frac{1}{2}} \ \ \ \ \ \textrm{(for $Y=Y_c$)}~.
\end{align}
Thus, from the definition of $\b_i$ (see \eqref{BET}), we get the value $\b_i=-1/2$.

Furthermore, when $M=M_c$ and $\bar Y_i=\bar Y_{ic}$, we obtain 
\begin{align}
\t X^i\sim \frac{1}{A^{(i)}_{001}}y_i^{-\frac{1}{2}} \ \ \ \ \ \textrm{(for $M=M_c$ and $\bar Y_i=\bar Y_{ic}$)}~.
\end{align}
In that case, 
\begin{align}
\eta_{Y_i}\sim y_i^{-\frac{1}{2}} \ \ \ \ \ \textrm{(for $M=M_c$ and $\bar Y_i=\bar Y_{ic}$)}~.
\end{align}
Therefore from the definition of $\d_i$ in \eqref{DEL}, we get $\d_i=-2$.

The numerical values of critical exponents obtained so far is given in the following table

\begin{table}[ht]
\begin{center}
\caption{Values of first set of critical exponents}

\label{table1}
~\\
\begin{tabular}{|c|c|c|c|c|c|c}
\hline \hline
$\a$ & $\phi_i$ & $\gamma_i$ & $\sigma_i$ & $\beta_i$ & $\delta_i$\\ \hline 
$\frac{3}{2}$ & $\frac{3}{2}$ & $\frac{3}{2}$ & $\frac{3}{2}$ & $-\frac{1}{2}$ & $-2$\\ \hline
 \hline
\end{tabular}
\end{center}
\end{table}
One can easily check above exponents satisfy the following scaling laws of ``first kind''.
\begin{align}
\a +2\b+\g=2~, \label{SCAL1}
\\
\b(\d-1)=\g~, \label{SCAL2}
\\
\phi(\b+\g)=\a~. \label{SCAL3}
\end{align}
The same values of the critical exponents were obtained earlier in \cite{Kaburakigrg, Cai:1996df} considering specific form of metrics. On the contrary, here we obtained those without the explicit information of the black hole spacetime by taking into account two inputs: (a) the black holes we considered here belong to the class, which exhibit extremal phase transition and (b) those black holes satisfy the first law of black hole mechanics. {\it This shows the universality of this type of critical phenomenon}.   

Apart from these critical exponents which were obtained above, there are a few others which are studied in the context of the extremal criticality. In the following, we shall discuss those critical exponents and shall obtain their values in a general manner. 
Near the critical point, the asymptotic form of the two point correlation function for large $r$ is defined by \cite{book}, 
\begin{align}
G(r)\sim \frac{e^{ (-r/\xi)}}{r^{d-2-\eta}}~. \label{CORR}
\end{align}
 Here, $\eta$ is called as the Fisher's exponent, $d$ is the effective spatial dimension and $\xi$ is called the correlation length. Near the critial point, the behaviour of $\xi$ is given as 
 \begin{eqnarray}
&& \xi\sim m^{-\nu}\ \ \ \ \ (\textrm{for all}~ Y_i=Y_{ic})~; \label{NU}
 \\
&& \xi\sim y_i^{-\mu_i}\ \ \ \ \ (\textrm{for $M=M_c$ and $\bar Y_i=\bar Y_{ic}$})~. \label{MU}
 \end{eqnarray}
 In the theory of quantum gravity, we do not have much knowledge about the two point correlation function defined in \eqref{CORR}. However, for extremal  Reissner-Nordstrom black hole, the inverse of the surface gravity is argued to play the role of the correlation length\cite{Traschen:1994md}. This result also holds for BTZ black hole \cite{Lifschytz:1993eb, Ichinose:1994rg, Cai:1996df} and black $p$-branes \cite{Cai:1997cs, Cai:1997cv}. If we assume, this to be true in the presence of multiple charges in arbitrary dimensions, we get $\xi\sim 1/\kappa\sim 1/T$.  Using \eqref{kl}, we can further conclude $\xi\sim (\p M/\p s)_{Y_i}^{-1}$. Therefore, from \eqref{MEXP}, the leading order contribution gives
 \begin{align}
 \xi\sim m^{-\frac{1}{2}} \ \ \ \ \ (\textrm{for all}~ Y_i=Y_{ic})~.
 \end{align}
 From the definition of $\nu$ in \eqref{NU}, we get the value $\nu=1/2$~. Now, when $M$ and all $Y$ are at their critical values except the $i^{th}$ charge $Y_i$, we obtain from \eqref{MEXP}
 \begin{align}
 \xi\sim y_i^{-\frac{1}{2}} \ \ \ \ \ (\textrm{for $M=M_c$ and $\bar Y_i=\bar Y_{ic}$})~.
 \end{align}
  Therefore, from \eqref{MU} we see that all $\mu_i$'s are the same and $\mu_i=\mu=1/2$.
  
  Now, these critical exponents are supposed to satisfy the scaling laws of ``second kind'', which are given by \cite{Stanley1, book},
  \begin{align}
  \nu(2-\eta)=\gamma~, \label{SCAL4}
  \\
  \nu d=2-\a~, \label{SCA5}
  \\
  \mu(\b+\g)=\nu~.
  \end{align}
  Using the obtained value of $\a$, $\b$, $\g$, $\mu$ and $\nu$ in the scaling law of the second kind, we get the value of remaining critical exponent $\eta$ and effective spacetime dimension $d$ . These are $\eta=-1$ and $d=1$~. Following table shows these values of exponents.
  
  \begin{table}[ht]
\begin{center}
\caption{Values of remaining critical exponents}

\label{table2}
~\\
\begin{tabular}{|c|c|c|c|c|c|c}
\hline \hline
$\nu$ & $\mu_i$ & $\eta$  \\ \hline 
$\frac{1}{2}$ & $\frac{1}{2}$ & $-1$ \\ \hline
 \hline
\end{tabular}
\end{center}
\end{table}
Remember, in the above analysis we have assumed that the correlation length is given by the inverse of the surface gravity. This has been checked and accepted for several instances \cite{Traschen:1994md, Lifschytz:1993eb, Ichinose:1994rg, Cai:1996df, Cai:1997cs, Cai:1997cv}. However, we are not sure if this is true in general. Therefore, it would be interesting if the same conclusion can be drawn from a general argument. For the time being, we leave that analysis for future. 
  
%--------------------------------------------------------------------------------------------

%###########################################################################################
\section{\label{sec4}GTD in extremal phase transition}
The concepts of differential geometry is used in thermodynamics for a long time. The underlying motivation to pursue in this direction is to study various thermodynamic phenomena in terms of the geometric properties of the phase space of the system. For non-extremal black holes, there are two major approaches of studying the phase transition of black hole-- one approach deals with the divergence of heat capacity and inverse of isothermal compressibility  \cite{Banerjee:2011cz, Banerjee:2012zm, Majhi:2012fz, Lala:2012jp, Ma:2014tka, Azreg-Ainou:2014gja, Liu:2013koa, Mandal:2016anc}. The other approach \cite{Kubiznak:2012wp, Kubiznak:2016qmn, Majhi:2016txt, Bhattacharya:2017nru} is for the black holes in the AdS background, in which the cosmological constant is treated as the thermodynamic pressure. The latter approach exactly resembles the phase transition of the van der Waals fluid system. It must be mentioned that both these phase transitions have been studied extensively under the light of the GTD \cite{Banerjee:2016nse, Bhattacharya:2017hfj, Dehyadegari:2018pkb}. Here people have formulated thermogeometrical metrics in the thermodynamic phase space of black hole and have shown that the corresponding Ricci-scalar diverges at the phase transition point.

 In this section, we incorporate those ideas to study the extremal phase transition. Here, we comment that there are several ways to formulate the thermogeometrical metric. First Weinhold \cite{WEINHOLD} introduced a metric, the components of which are given by the Hessian of the internal thermodynamic energy. Later, Ruppeiner \cite{RUPP, Ruppeiner:1995zz} introduced another metric, which is defined as the negative of the Hessian of the entropy, and is conformal to the Weinhold metric with the conformal factor being the inverse temperature. Later, Quevedo \cite{Quevedo:2006xk, Quevedo:2007mj, Quevedo:2008xn, Quevedo:2008ry, Alvarez:2008wa, Quevedo:2011np, Quevedo:2017tgz, Quevedo:2016cge} came up with the idea of defining the thermogeometrical metric in a Legendre-invariant way. 
 
 In our general procedure of analyzing the extremal phase transition, we study the behaviour of Ricci-scalar near the critical point for all these metrics.

\subsection{The Weinhold metric}
To write the Weinhold metric, one has to write mass (which plays the role of internal energy) as the function of entropy and the charges i.e., $M\equiv M (S, Y_i)$. Now for the sake of simplicity we consider the dependence of mass on a particular charge $Y$ and keep all other charges fixed. Therefore the first law of thermodynamics is written as,
\begin{align}
dM=TdS+XdY~. \label{1st Law}
\end{align}
Here $T=(\p M/\p S)_Y$ and $X=(\p M/\p Y)_S$. 

Now the Weinhold metric is given by,
\begin{align}
ds_W^2=\frac{\p^2 M}{\p x_i\p x_j}dx_idx_j\ \ \ \ \ \ \ \ \ \ \ \ \ \ \ \ \{x_1=S,\ x_2=Y\}~. \label{WEIN}
\end{align}
The expanded form of the Weinhold metric is 
\begin{align}
ds^2_W=-f(S,Y)dS^2+g(S,Y)dY^2+2h (S,Y)dSdY~, \label{WEINEXT}
\end{align}
where $f(S, Y)=-M_{SS}$, $g(S, Y)=M_{YY}$ and $h(S, Y)= M_{SY}=M_{YS}$~. The Ricci scalar corresponding to the Weinhold metric \eqref{WEINEXT} is given by,
\begin{align}
\no
R_{(W)}=\frac{1}{2(fg+h^2)^2}\Big[f(f_Y g_{Y}-g_S^2+2g_Yh_S)+g\Big\{f_Y^2+f_S(2h_Y-g_S)-2f(f_{YY}+h_{SY}-g_{SS})\Big\} 
\\ 
+h\Big\{-g_Yf_S+f_Y(2h_Y+g_S)+4h_Yh_S-2g_Sh_S-2h(f_{YY}+2h_{SY}-g_{SS})\Big\}\Big] ~,\label{RICCIWEIN}
\end{align}
where $f_J=\p f/\p J$ and so on. Now, from the expansion of $M$ (given in \eqref{MEXP}) we can conclude that $f$, $g$, $h$ and their derivatives are finite. Therefore, the Ricci scalar of the Weinhold metric is a finite quantity near the critical point.

%--------------------------------------------------------------------------------------------

%###########################################################################################
\subsection{The Ruppeiner metric}
We first write the first law of thermodynamics\eqref{1st Law} as $dS=\b dM-\t X dY$. In this form, the conjugate quantities are taken as $\b=(\p S/\p M)_Y$ and $\t X=-(\p S/\p Y)_M$. Now, the Ruppeiner metric is defined as
\begin{align}
ds^2_R=-\frac{\p^2S}{\p x_i'\p x_j'}dx_i'dx_j'\ \ \ \ \ \ \ \ \ \ \ \ \ \ \ \ \{x_1'=M,\ x_2'=Y\}~. \label{RUP}
\end{align}
Here, $g_{11}=-S_{MM}$, $g_{22}=-S_{YY}$ and $g_{12}=g_{21}=-S_{MY}$~. It implies that the expansion of the Ruppeiner metric is 
\begin{align}
ds^2_R=-f'(M,Y)dM^2+g'(M,Y)dY^2+2h' (M,Y)dMdY~, \label{RUPEXT}
\end{align}
where $f'=S_{MM}$, $g'=-S_{YY}$ and $h'=-S_{MY}$~.
The Ricci scalar of the metric \eqref{RUPEXT} is found to be, 
\begin{eqnarray}
&& R_{(R)}=\frac{1}{2(f'g'+h'^2)^2}\Big[f'(f_Y' g_{Y}'-g_M'^{2}+2g_Y'h_M')\no 
\\
&&+g'\Big\{f_Y'^2+f_M'(2h_Y'-g_M')-2f'(f_{YY}'+h_{MY}'-g_{MM}')\Big\}\no 
\\
&&+h'\Big\{-g_Y'f_M'+f_Y'(2h_Y'+g_M')+4h_Y'h_M'-2g_M'h_M'-2h'(f_{YY}'+2h_{MY}'-g_{MM}')\Big\}\Big] \label{RICCIRUP}
\end{eqnarray}
Now, we have to calculate each term of the Ricci scalar of \eqref{RICCIRUP} to see its dependence on $s$. To do that, we find out the leading order contribution of $f'$, $g'$ and their derivatives.
%\fbox{\begin{minipage}{44em}
% Using the mass ($m$) order of $T$ (or $\b$), $C_Y$ and $I_M$ which was obtained earlier, one can show that $f'\sim m^{-\frac{3}{2}}$, $f_{x_i'}'\sim \frac{\b^2}{C_J^2} \sim m^{-2}$, $f_{x_i'x_j'}'\sim \frac{\b^2}{C_J^3}\sim m^{-\frac{5}{2}}$ and similarly $g'\sim m^{-\frac{3}{2}}$, $g_{x_i'}'\sim m^{-\frac{5}{2}}$ and $g_{x_i'x_j'}'\sim m^{-\frac{7}{2}}$. We put all these in Eq. \eqref{RICCIRUP}. We find that the second and the ninth term of Ricci scalar diverges in the order of $m^{-\frac{1}{2}}$. The other terms are the order of $m^F$, where $F$ either zero or positive number (like $\frac{1}{2}, 1, \frac{3}{2}, 2\ etc.$). Also, It can be checked that the two diverging terms does not cancel each other. Therefore, we can conclude that the Ricci scalar of the Ruppeiner metric diverges and the divergence of the Ricci scalar is given as
%\begin{align}
%R_{(R)}\sim m^{-\frac{1}{2}}~.
%\end{align}
%\end{minipage}}
%\vskip 3mm
%Let us do it in a different way. 
From \eqref{P2SPM2} we see that $f'=-(\p^2S/\p M^2)_{Y}\sim 1/s^3$. Therefore, $f_M'=(\p f'/\p M)_Y\sim (1/s^4)(\p s/\p M)_Y$. Using \eqref{kl}, one obtains $f_M'\sim s^{-5}$. In a similar way, $f_{MM}'\sim s^{-7}$. Now,  $f_Y'=(\p f'/\p Y)_M\sim (1/s^4)(\p s/\p Y)_M$. Again, using \eqref{PYPS} one gets $f_Y'\sim s^{-5}$. The same arguments yield $f_{YY}'\sim s^{-7}$ and $f_{MY}'=f_{YM}'\sim s^{-7}$~. Following the same procedure, one similarly obtains $g'\sim s^{-3}$, $g'_{x_i'}\sim s^{-5}$ and $g'_{x_i'x_j'}\sim s^{-7}$. Also, $h'\sim s^{-3}$, $h'_{x_i'}\sim s^{-5}$ and $h'_{x_i'x_j'}\sim s^{-7}$. As a result, we see that the denominator goes as $\sim s^{-12}$ and each term in the numerator goes as $\sim s^{-13}$. Therefore, the Ricci scalar diverges as 
\begin{align}
R_{(R)}\sim s^{-1}~.
\end{align}
The property of the Ruppeiner metric has also been studied in a different way \cite{Cai:1998ep, Wei:2009zzf} while studying the extremal phase transition of BTZ black holes. It has been there argued that the Ruppeiner metric should diverge as $R_{(R)}\sim \xi^{d}$. Since, in our case $\xi\sim s^{-1}$ near the critical point, we obtain $R_{(R)}\sim \xi^{1}$. Therefore, we can again conclude that the effective spatial dimension $d=1$ for any extremal black hole, which is in agreement with the claim of the recent papers \cite{Horowitz:1996ac, Ghosh:1998cc}. Thus, from the thermogeometric approach, we can again generally prove that the effective spatial dimension of an extremal black hole is one.
%--------------------------------------------------------------------------------------------

%###########################################################################################
\subsection{Legendre invariant metric}
Above two thermogeometrical metrics, namely the Wienhold and the Ruppeiner metric are not Legendre-invariant. Moreover in some cases, conclusions derived from the Weinhold metric and the Ruppeiner metric are not consistent with each other. Later Quevedo {\it et al.} claimed that those inconsistencies appear because these metrics are not Legendre-invariant and hence they came up with Legendre invariant metric formalism \cite{Quevedo:2006xk, Quevedo:2007mj, Quevedo:2008xn, Quevedo:2008ry, Alvarez:2008wa, Quevedo:2011np, Quevedo:2017tgz, Quevedo:2016cge}. In the following, we discuss two types of Legendre invariant thermogeometrical metric. One of them (Quevedo metric: 1) is mostly used as a Legendre-invariant metric. Here, we see that the Ricci scalar of the first type of the Legendre-invariant metric is a finite quantity at the critical point. So we discuss another type of Legendre-invariant metric (Quevedo metric: 2). The second metric is not that familiar but we see that the Ricci scalar corresponding to this metric vanishes. The formalism which we adopt here was originally developed by Hermann \cite{Hermann} and Mrugala \cite{Mrugala1, Mrugala2}, which was later followed extensively by Quevedo.

\subsubsection{Quevedo metric: 1}
We define a thermodynamic phase space $\mathcal{T}$ with coordinates $\mathcal{Z}^A=\{S, q^a, p^a\}$ where $q^a=\{M, Y\}$ are the variables and $p^a=\{S_M=\b, S_Y=-\t X=-\b X\}$ are the conjugate variables. Therefore, in the entropy representation, the fundamental one form in $\mathcal{T^*}$ (where, $\mathcal{T^*}$ is the cotangent  space of $\mathcal{T}$) is given by,
\begin{align}
\Theta_S=dS-\b dM+\t X dY~, \label{FUNDA}
\end{align}
which is invariant under the Legendre transformation 
\begin{align}
M(q)=\t M(\t q)-\d_{ab}\t{q}^a\t{p}^b \label{LEGENDRE}
\\
\no 
\textrm{with}\ \ \ \ \ \ \ \ \ \ 
\ \ q^a=-\t p^a \ \ \textrm{and}\ \ p^a=\t q^a~.
\end{align}
Now, following Quevedo's formalism, one possible form of the Legendre invariant thermogeometrical metric (on $\mathcal{T}$) is (Eq. (39) of \cite{Quevedo:2006xk})
\begin{align}
G_1=\Theta_S^2+ (\b M+\t XY)(d\b dM+dYd\t X).
\end{align}
Expanding the conjugate quantities ($\b$ and $\t X$) as a function of the variables ($M$ and $Y$), one finds the the expression of $G_1$ in the space of equilibrium ($\Theta_S=0$) as
\begin{align}
G_1=-f_1(M, Y)dM^2+g_1(M, Y) dY^2~, \label{METLEG}
\end{align}
where $f_1(M, Y)=-(\b M+\t XY)S_{MM}$ and $g_1(M, Y)=-(\b M+\t XY)S_{YY}$~. The Ricci scalar of the metric \eqref{METLEG} is given by,
\begin{align}
R_1=\frac{1}{2(f_1g_1)^2}\Big[f_1(f_{1Y} g_{1Y}-g_{1M}^{2})+g_1\Big\{f_{1Y}^2-f_{1M}g_{1M}-2f_1(f_{1YY}-g_{1MM})\Big\}\Big] ~. \label{RICCI1}
\end{align}
Again, we check the order of each term in the Ricci scalar. $f_1\sim \b S_{MM}\sim (\p S/\p M)_Y(\p^2 S/\p M^2)_Y$. This implies $f_1\sim s^{-4}$. Similarly $g_1\sim s^{-4}$. Following the same procedure as was done in the Ruppeiner case, we obtain $f_{1x_i}\sim s^{-6}$,  $g_{1x_i}\sim s^{-6}$, $f_{1x_i x_j}\sim s^{-8}$ and $g_{1x_ix_j}\sim s^{-8}$. Therefore, we see that the denominator goes as $\sim s^{-16}$ and the numerator also goes as $\sim s^{-16}$. Therefore, the Ricci scalar is finite in this case.
\subsubsection{Quevedo metric: 2}
As the choice of Legendre invariant metric is not unique, we can formulate other Legendre invariant metric. Following Quevedo's formalism (Eq. (37) of \cite{Quevedo:2006xk}) we see 
\begin{align}
G_2=\Theta_S^2+c_1\b M d\b dM+c_2 \t X Yd\t XdY+d\b^2+dM^2+d\t X ^2+dY^2 \label{METR2}
\end{align}
is Legendre invariant for any value of the real constants $c_1$ and $c_2$. For the simplicity of calculation, we take $c_1=c_2=1$~. Now using $d\b=S_{MM}dM+S_{MY}dY$ and $d\t X=-S_{YM}dM-S_{YY}dY$ in \eqref{METR2} we get, in equilibrium space
\begin{align}
G_2=-f_2(M,Y)dM^2+g_2(M,Y)dY^2+2h_2 (M,Y)dMdY~,
\end{align}
where $f_2=-[1+\b M S_{MM}+S_{MM}^2+S_{MY}^2]$, $g_2=1-\t X YS_{YY}+S_{YY}^2+S_{MY}^2$ and $h_2=\frac{1}{2}(\b M-\t X Y) S_{MY}+S_{MM}S_{MY}+S_{YM}S_{YY}$. Thus the Ricci scalar is given by, 
\begin{eqnarray}
&&R_2=\frac{1}{2(f_2g_2+h_2^2)^2}\Big[f_2(f_{2Y} g_{2Y}-g_{2M}^{2}+2g_{2Y}h_{2M})
\no 
\\
&&+g_2\Big\{f_{2Y}^2+f_{2M}(2h_{2Y}-g_{2M})-2f_2(f_{2YY}+h_{2MY}-g_{2MM})\Big\}
\\
\no 
&&+h_2\Big\{-g_{2Y}f_{2M}+f_{2Y}(2h_{2Y}+g_{2M})+4h_{2Y}h_{2M}-2g_{2M}h_{2M}-2h_2(f_{2YY}+2h_{2MY}-g_{2MM})\Big\}\Big] \label{RICCI2} 
\end{eqnarray}
Now, $f_2=\mathcal{O}(s^0)+\mathcal{O}(s^{-4})+\mathcal{O}(s^{-6})$. The leading order contribution near the critical point will be $f_{2}\sim s^{-6}$. As a result, $f_{2x_i}\sim s^{-8}$ and $f_{2x_ix_j}\sim s^{-10}$~. Leading order contributions of $g_{2}$ and $h_2$ are same as $f_2$. Therefore, the denominator goes as $\sim s^{-24}$ and the numerator goes as $\sim s^{-22}$. As a result, 
\begin{align}
R_2\sim s^2~.
\end{align}
Consequently, we see that the Ricci-scalar vanishes near the critical point.

In this section, we have studied the behaviour of the Ricci-scalar for different thermogeometrical metrics and have shown that the Ricci-scalar of the Ruppeiner metric diverges at the extremal limit. On the contrary, the Ricci-scalar of other thermogeometrical metrics remains finite (or vanishes) at that point. Therefore, we conclude that the extremal phase transition shows the behaviour of the second order phase transition not only in the specific ensemble of thermodynamics (\ie the microcanonical ensemble), but also for a specific thermogeometric manifold as well (the Ruppeiner one). Note that the Legendre-invariant thermogeometrical metrics, which are mostly used nowadays, cannot confirm the second order phase transition in the present case. A plausible explanation to that might be as follows. Remember that the Legendre-invariant metrics are constructed on the line of arguments that a proper thermogeometrical metric should be Legendre invariant as the thermodynamic features are invariant in all ensembles. Since one thermodynamic potential, by which an ensemble is characterized, is connected to the same in the other ensemble by the Legendre transformation, the entire thermodynamic description is invariant due to the Legendre transformation, which should reflect on the thermogeometrical metric. However, as we have noticed in the present case, the identification of the non-extremal to extremal transformation with the second order phase transition is valid only in the microcanonical ensemble. As a result, the present thermodynamic description is not invariant across all ensembles. Therefore, the use of a Legendre-invariant metric might not be suitable in this case. Nonetheless, we have checked the behaviour of the Ricci-scalar of all the thermogeometrical metrics which are popular in GTD and from that analysis we found that the Ruppeiner metric is the ideal one for the thermogeometric description of the extremal phase transition. {\it Interestingly, here entropy $S$ plays the central role both in microcanonical ensemble ($S$ is chosen as the Massieu function) and in Ruppeiner geometrical description (the metric is constructed by considering $S$ as the thermodynamic potential)}. 

%-----------------------------------------------------------------------------------------------
%%%%%%%%%%%%%%%%%%%%%%%%%%%%%%%%%%%%%%%%%%%%%%%%%%%%%%%%%%%%%%%%%%%%%%%%%%%%%%%%%%%%%%%%%%%%%%%%
\section{Conclusions}
In this work, we have studied the extremal phase transition of black hole in a general framework. There are several works \cite{Kaburakigrg, curir1, curir2, Pavon:1988in, Pavon:1991kh, Cai:1996df, Cai:1998ep, Wei:2009zzf, Cai:1997nb, Ma:2013eaa} to show that the extremal phase transition is a second order phase transition. These earlier works were done case by case for a particular spacetime and dimension. The obtained results in different spacetimes (such as the critical exponents, scaling laws \etc) are in accordance with each other and strongly suggest that there must be a metric independent way to establish those earlier results. This has been the major motivation for this work.

We have proved that the transformation of the black hole from a non-extremal to an extremal one is a second order phase transition. For that, we have calculated the second order moments of fluctuations in different ensembles and have shown that those moments diverge for a black hole in microcanonical ensemble, which is a sign of a second order phase transition as per the prescription of Pav$\acute{\textrm{o}}$n and Rub$\acute{\textrm{i}}$ \cite{Pavon:1988in, Pavon:1991kh}. Afterwards, we have generally obtained the critical exponents for this phase transition and have shown that the critical exponents satisfy the scaling laws. While proving those results, we have not accounted any particular spacetime, which implies our results are valid for all the black hole spacetimes which become extremal at certain limit. Thus, the universality of results, which were predicted by earlier works, is proved by our analysis and hence from now on one need not check the critical behaviour case by case. 

Finally, we have extended our analysis to GTD, which is a recent formalism to describe the phase transition geometrically. We have shown that the extremal critical point of black holes can be identified as a particular point where the Ricci scalar corresponding to the Ruppeiner metric diverges. In addition, we have also shown that the Ricci scalar of the Weinhold metric and of one type of Legendre-invariant metric (Quevedo metric: 1) is a finite quantity and does not show any special behaviour. In another Legendre invariant metric (Quevedo metric: 2), the Ricci-scalar vanishes on the critical point. In this analysis we observed that extremal phase transition is properly explained in microcanonical ensemble and by Ruppeiner geometry. Note that in both the descriptions, entropy plays the central role: $S$ acts as Massieu function in microcanosical ensemble and thermodynamical potential in GTD. At this moment, the actual reason for this is not known to us; hope we shall be able to find the precise reason in future.  

Thus our paper covers different thermodynamics aspects of extremal black hole. Other previous works in this field confined their analysis to specific cases and hence can not explain questions regarding universality. The novelty of our work is, it is very general and does not require any specific metric. In this sense our paper unifies all other work on extremal phase transition in an elegnant manner. At last we shall conclude by making the following comments on our observations we made here on the extremal phase transition.

 In this work, we have examined whether any phase transition occurs during the transition of a black hole from a non-extremal to an extremal one. For that, in our general framework (i.e. without using the explicit expression for black hole metric), we have taken the help of the fluctuation theory. It has been observed that the presence of a second order phase transition naturally occurs only in the microcanonical ensemble, while the other ensembles (canonical and grand canonical) fail to show that. This has also been observed earlier in several case by case studies (i.e. explicitly using the black hole metric expression) \cite{Kaburakigrg, Cai:1996df, Cai:1998ep, Wei:2009zzf}. The possible reasons for that can be stated as follows. 
In this context, let us first mention why not all ensembles agree upon the same result in the fluctuation theory. Usually, we see that the mean values of different thermodynamic quantities are the same in different ensembles for a given system in equilibrium. However, it must be noted that the different ensembles predict different fluctuations of a thermodynamic parameter around its equilibrium value \cite{Kaburakigrg}. In other words, average value of thermodynamic quantities are same in all ensembles, but fluctuations are not.  Thus, the usual notion of the equivalence of the different ensembles can break down while investigating the physics with the help of fluctuations in the macroscopic parameters. We also have observed the same in the present analysis as well. Only in the microcanonical ensemble all the second order moments of the relevant quantities are divergent and implies the presence of the critical point. While in other ensembles (canonical and grand canonical) one cannot confirm the presence of the critical point at $T=0$ as all the second order fluctuation modes do not diverge in those cases.

Let us now understand why the microcanonical ensemble appears to be so special in this case. Remember, in several cases of black hole thermodynamics, one particular ensemble (specially the microcanonical ensemble) can be more preferred then the other ensembles. For example, the microcanonical ensemble is the most suitable one for the discussion of the fluctuations of stellar mass or more massive black holes. This is because the time scale of particle exchange is much larger than the present age of the universe in such cases \cite{Kaburakigrg}, which means the black hole hardly exchange any particle with the environment. On the contrary, if the black hole is small, more particle exchange can take place and the grand canonical ensemble becomes more suitable for the thermodynamic description. Another example is that the microcanonical ensemble is the proper ensemble for the thermodynamic description of the microscopic black holes which are not in equilibrium, such as the radiating black holes \cite{Casadio:2011pd}. This example is particularly important in this case because we have accounted the temperature and entropy of the black holes, which is obtained only when one considers the quantum (microscopic) effect in the theory. Thus, it can be concluded that in certain cases, one particular ensemble can be more favourable then the others in black hole thermodynamics. From that line of argument, it can be said that the microcanonical ensemble can be the appropriate or a proper ensemble  for the thermodynamic description of extremal phase transition of black holes.

 Later from our thermogeometric analysis, we have found that the divergence of the Ricci-scalar at the critical point occurs only for the Ruppeiner metric, whereas the scalar curvature is either finite or vanishing for the Weinhold and Quevedo (I and II) metrics. Firstly, we mention why the Ruppeiner metric is unique in this study. It would be interesting to note that the Ruppeiner metric is the Hessian of the Massieu function of the microcanonical ensemble (the entropy), which, as we have observed earlier, can be regarded as the proper ensemble for the thermodynamic description of the extremal phase transition of black holes. From that viewpoint, the Ruppeiner metric is special in this case, in spite of the fact that this metric is not formulated in a Legendre invariant way. 

Now, we mention why the Legendre-invariant formalism by Quevedo has not been able to reflect the extremal phase transition through the divergence of corresponding Ricci scalar. We have already seen, our analysis can predict the criticality only in the microcanonical ensemble. On the other hand, the Legendre-invariant way of defining thermogeometrical metric implies the result should be valid in all the ensembles. Since there is a pre-existing in-equivalence among the ensembles in the extremal phase transition, it is not surprising that the Legendre-invariant formulation is not suitable in the present case. Again, the root lies in the fact that we are looking at the average value (here it is Ricci scalar), not on the moments of the fluctuations (like $<\delta R \delta R>$) which can be different in different Legendre invariant metrics. Having the feel that the fluctuations in Ricci scalar can be good quantity in explaining the extremal phase transition in the context of thermogeometric study of phase transition, we calculated $<\delta R \delta R>$ for both the Quevedo metrics. The details of this is presented in Appendix \ref{AppC}. 
We found that the moments of fluctuation of the Ricci-scalar diverges at the critical point for Quevedo-I metric, which is mostly used in the thermogeometric description. {\it Thus, it can be conjectured that instead of the Ricci-scalar, from the study of the fluctuation of the Ricci-scalar the presence of the criticality can be well determined}.

 % As we have emphasized several times, the results are very general and are not restricted for any particular spacetime, unlike all other previous works in this regard. Therefore, the novelty of this work lies in its generality and the universality of the results which are obtained in this analysis. Black hole phase transition is an active area of research work for several years. We hope this work will help people to understand the subject more evidently. Also, we expect to clarify some more issues in this regard and report soon.
%-------------------------------------------------------------------------------------------------
%%%%%%%%%%%%%%%%%%%%%%%%%%%%%%%%%%%%%%%%%%%%%%%%%%%%%%%%%%%%%%%%%%%%%%%%%%%%%%%%%%%%%%%%%%%%%%%%%%
\section*{Appendix}
\appendix
\section{Obtaining the values of $(\p^2 \Phi_2/\p \b^2)_{Y_i}$ and $(\p^2 \Phi_2/\p Y_i^2)_{\b, \bar Y_i}$} \label{APPENCANO}
We take the canonical ensemble in which Helmholtz function is $F\equiv F(T, Y_i)$~. Equivalently one can write $T\equiv T(F, Y_i)$~. As we have done earlier, we expand $T$ around the critical point $T_c=0$ which yields,
 \begin{eqnarray}
\no
&& T=b_{10}f+b_{20}f^2+b_{30}f^3+b_{40}f^4+... 
\\
\no 
&& \ \ \ \ \ \ +b_{01}^{(1)}y_1+b_{02}^{(1)}y_1^2+b_{03}^{(1)}y_1^3+b_{04}^{(1)}y_1^4+...\\
\no 
&& \ \ \ \ \ \ +b_{01}^{(2)}y_2+b_{02}^{(2)}y_2^2+b_{03}^{(2)}y_2^3+b_{04}^{(2)}y_2^4+...\\ 
&& \ \ \ \ \ \ \ \ . . .  + b_{ij}^{(k)}f^iy_k^j~,\label{TEXP}
 \end{eqnarray}
where, $f=F-F_c$ and so on. In the above expansion, we have used $T_c=0$. Now keeping terms upto first order we get,
 \begin{align}
\frac{\p F}{\p T}\Big|_{Y_i}= \frac{\p T}{\p F}\Big|_{Y_i}^{-1}\sim \frac{1}{B_{00}+B_{10}f+B^{(i)}_{11}y_i} ~. \label{DFDTY}
 \end{align}
 and 
 \begin{align}
 \frac{\p^2 F}{\p T^2}\Big|_{Y_i}\sim\frac{\p}{\p T}\Big(\frac{1}{B_{00}+B_{10}f}\Big)\Big|_{Y_i}\sim \frac{1}{(B_{00}+B_{10}f)^2+B^{(i)}_{11}y_i}\frac{\p F}{\p T}\Big|_{Y_i}\sim \frac{1}{(B_{00}+B_{10}f+B^{(i)}_{11}y_i)^3}~.
 \end{align}
 It implies that $(\p^2 F/\p T^2)_{Y_i}$ is a non-zero finite quantity at the critical point and near that point, it goes as $(\p^2 F/\p T^2)_{Y_i}\sim B_{00}^{-3}$~.

Now to obtain $(\p^2 F/\p Y_i^2)_{T, \bar Y_i}$, we expand $Y_i$ near the critical point as a function of $T$, $F$ and $\bar Y_i$~. This is
\begin{eqnarray}
\no
&& Y_i={Y_i}_c+b_{100}f+b_{200}f^2+b_{300}f^3+b_{400}f^4+... 
\\
&&\ \ \ \ \ \ \ \ \no b_{010}T+b_{020}T^2+b_{030}T^3+b_{040}T^4+...
\\ 
&& \ \ \ \ \ \ \ \ . . . + b_{jkl}f^jT^k\bar y_i^l~.\label{YEXP2}
 \end{eqnarray}
 In the above equation, we have used $T_c=0$. Again, adopting the similar method as earlier, it can be shown straightforwardly that $(\p^2 F/\p Y_i^2)_{T, \bar Y_i}$ is also a non-zero finite quantity at the critical point.
 
 As $\Phi_2=-\b F$, one can straightforwardly obtain $(\p^2 \Phi_2/\p \b^2)_{Y_i}=-T^3(\p^2 F/\p T^2)_{Y_i}$. Therefore at the critical point, $(\p^2 \Phi_2/\p \b^2)_{Y_i}$ vanishes as 
 \begin{align}
 \Big(\frac{\p^2 \Phi_2}{\p \b^2}\Big)_{Y_i}\sim T^3~.
 \end{align}
Again, $(\p^2 \Phi_2/\p Y_i^2)_{\b, \bar Y_i}=\b(\p^2 F/\p Y_i^2)_{T, \bar Y_i}$. Therefore, at the critical point, $(\p^2 \Phi_2/\p Y_i^2)_{\b, \bar Y_i}$ diverges as
\begin{align}
\Big(\frac{\p^2 \Phi_2}{\p Y_i^2}\Big)_{\b, \bar Y_i}\sim T^{-1}~.
\end{align}

%---------------------------------------------------------------------------------------------------

\section{Obtaining the values of $(\p^2 \Phi_3/\p \b^2)_{\t X^i}$ and $(\p^2 \Phi_3/\p \t {X^i}^2)_{\b, \bar{\t X}^i}$} \label{APPENGRAND}
Let us take the Gibbs free energy $G\equiv G(T, X^i)$. Alternatively temperature is written as $T\equiv T(G, X^i)$~. Now expanding $T$ near the critical point, as we have done earlier, it can be shown that $(\p^2 G/\p T^2)_{X_i}$ is a non-zero finite quantity. Similarly, expanding $X^i$ in terms of $T$, $G$ and $\bar{X}^i$, one finds that $(\p^2 G/\p {X^i}^2)_{T, \bar{X}^i}$ is also a non-zero finite quantity. Now, as $\Phi_3=-\b G$, we obtain $(\p^2 \Phi_3/\p \b^2)_{\t X^i}=-T^3(\p^2 G/\p T^2)_{X_i}$. Therefore, we conclude that near the critical point $(\p^2 \Phi_3/\p \b^2)_{\t X^i}$ vanishes as 
\begin{align}
\Big(\frac{\p^2 \Phi_3}{\p \b^2}\Big)_{\t X^i}\sim T^3~.
\end{align}
Now using $\t X^i=\b X^i$, one can show $(\p^2 \Phi_3/\p \t {X^i}^2)_{\b, \bar{\t X}^i}=T(\p^2 G/\p {X^i}^2)_{T, \bar{X}^i}~.$ Hence, near the critical point, $(\p^2 \Phi_3/\p \t {X^i}^2)_{\b, \bar{\t X}^i}$ vanishes as 
\begin{align}
\Big(\frac{\p^2 \Phi_3}{\p \t {X^i}^2}\Big)_{\b, \bar{\t X}^i}\sim T~.
\end{align}

%%%%%%%%%%%%%%%%%%%%%%%%%%%%%%%%%%%%%%%%%%%%%%%%%%%%%%%%%%%%%%%%%%%%%%%%%%%%%%%%%

%_------------------------------------------------------------------------------
\section{\label{AppC}Moments of fluctuations of Ricci scalar $\langle\d R \d R\rangle$ in Legendre invariant metrics}
\subsection{Quevedo-I metric}
The expression of the Ricci scalar for the metric Quevedo I is given in \eqref{RICCI1}. Let us now calculate the fluctuation of $R_1$. We obtain,
\begin{align}
\delta R_1 ={}& \frac{1}{2(f_1 g_1)^2}\Big[\delta f_1\Big\{f_{1Y}g_{1Y}-g_{1M}^2-2g_1f_{1YY}+2g_1g_{1MM}\Big\}+\delta g_1\Big\{f_{1Y}^2-f_{1M}g_{1M}- 2f_1f_{1YY}+2f_1g_{1MM}\Big\}\notag\\
&+\delta f_{1Y}\Big\{f_1g_{1Y}+2g_1f_{1Y}\Big\}+\delta f_{1M}\Big\{-g_1g_{1M}\Big\}+\delta g_{1Y}\Big\{f_1f_{1Y}\Big\}+\delta g_{1M}\Big\{-2f_1g_{1M}-g_1f_{1M}\Big\}\notag\\
&+\delta f_{1YY}\Big\{-2f_1g_1\Big\}+\delta g_{1MM}\Big\{2f_1g_1\big\}\Big]\notag\\
&+\delta f_1 \Big[-\frac{1}{f_1^3 g_1^2}\Big\{f_1\Big(f_{1Y}g_{1Y}-g_{1M}^2\Big)+g_1\Big\{f_{1Y}^2-f_{1M}g_{1M}-2f_1(f_{1YY}-g_{1MM})\Big\}\Big\}\Big]\notag\\
&+\delta g_1 \Big[-\frac{1}{f_1^2 g_1^3}\Big\{f_1\Big(f_{1Y}g_{1Y}-g_{1M}^2\Big)+g_1\Big\{f_{1Y}^2-f_{1M}g_{1M}-2f_1(f_{1YY}-g_{1MM})\Big\}\Big\}\Big]~. \label{DELRICCI}
\end{align}
First, let us concentrate on $\d f_1$, the expression of which is given as
\begin{align}
\delta f_1 =-(M \delta \b+ Y\delta \t X)S_{MM}-(\b M+\t XY)\delta (S_{MM})~.
\label{delf}
\end{align}
Note, while obtaining the above fluctuation, we have considered the control parameters ($M$, $Y$) to be fixed as we are concerned with the off-equilibrium variations and have accounted the variation of the conjugate quantities $\delta \b $ and $\delta \t X$ to be independent. Similarly one finds, 
\begin{align}
\delta g_1 = -\Big(M \delta \b + Y \delta \t X \Big)S_{YY} - \Big(\b M +X \t Y\Big) \delta(S_{YY})~.
\label{delg}
\end{align}
Our final aim, in this case, is to compute the moments of $\d R_1$, which will be very clumsy if we consider the whole expression of \eqref{DELRICCI}. Therefore, we consider term by term.  In $\langle\d R\d R\rangle$, we have several terms like $T_1= \langle\delta f_1 \delta f_1\rangle(f_{1Y}g_{1Y}-g^2_{1M}-2g_1f_{1YY}+2g_1 g_{1MM})^2/(4f_1^4 g_1^4) $, $T_2=\langle\delta f_1\delta g_1\rangle(f_{1Y}g_{1Y}-g^2_{1M}-2g_1f_{1YY}+2g_1 g_{1MM})(f^2_{1Y}-f_{1M}g_{1M}-2f_1f_{1YY} + 2f_1g_{1MM})/(4f_1^4 g_1^4)$, $T_3=\langle\delta f_1 \delta f_{1Y}\rangle(f_{1Y}g_{1Y}-g^2_{1M}-2g_1f_{1YY}+2g_1 g_{1MM})(f_1 g_{1Y}+ 2 g_1 f_{1Y})/(4f_1^4 g_1^4)$ and so on. Now concentrate on the following term:
\begin{align}
\Big<\delta f_1 \delta f_1\Big> ={}&\Big\{M^2 <(\delta \b)^2>+Y^2\Big<(\delta \t X)^2\Big>\Big\}S_{MM}^2+2S_{MM}(\b M + \t XY)\Big\{M\Big<\delta \b \delta (S_{MM})\Big> \notag\\
&+Y\Big<\delta \t X \delta (S_{MM})\Big>\Big\}+(\b M + \t XY)^2 \Big<\Big\{\delta (S_{MM})\Big\}^2\Big>.
\label{delfcorr}
\end{align}
From the equations \eqref{DELBDELB}, \eqref{DELXDELX1} and \eqref{P2SPM2} we see that $\Big<(\delta \b)^2\Big>$ and $\Big<(\delta \t X)^2\Big>$ diverge as $s^{-3}$.

For the present case, since we have not considered any particular spacetime, we are unaware of the expression of the entropy. So, we cannot definitely obtain the forms of the terms like $\d S_{MM}$, $\d S_{YY}$, $\d S_{MM}$ \etc~Therefore, it is hard to predict the order of the divergences of $\Big<(\delta \b \delta (S_{MM})\Big>$, $\Big<(\delta \t X \delta (S_{MM}))\Big>$ and $\Big<\Big\{\delta (S_{MM})\Big\}^2\Big>$. But the nature of the first term of the above at the critical point can be predicted in out present general approach. Using our earlier results $f_1 \sim s^{-4}$, $g_1 \sim s^{-4}$, $f_{1x_i} \sim s^{-6}$, $g_{1x_i} \sim s^{-6}$, $f_{1x_ix_j} \sim s^{-8}$ and $g_{1x_ix_j} \sim s^{-8}$, we obtain that the first term on the RHS of \eqref{delfcorr} diverges as $\sim s^{-9}$. Using the fact that $\langle\delta f_1 \delta f_1\rangle$ diverges as $s^{-9}$ near the critical point, we obtain $T_1$ diverges as $s^{-1}$. In a similar vein, the calculable or the known divergences in $\Big<\delta f_1 \delta g_1\Big>$ are of the order $\sim s^{-9}$. The same procedure yields the known divergences of the following correlators as 
\begin{eqnarray}
&&\Big<\delta f_1 \delta f_{1x_i}\Big> \sim s^{-11}; \,\,\,\
\Big<\delta f_1 \delta g_{1 x_i}\Big> \sim s^{-11}; \,\,\,\
\Big<\delta g_1 \delta f_{1 x_i}\Big> \sim s^{-11}; \,\,\
\Big<\delta g_1 \delta g_{1 x_i}\Big> \sim s^{-11}; 
\nonumber
\\
&&\Big<\delta f_{1x_i} \delta f_{1 x_j}\Big> \sim s^{-13}; \,\,\,\
\Big<\delta f_{1x_i} \delta g_{1 x_j}\Big> \sim s^{-13}; \,\,\,\
\Big<\delta g_{1x_i} \delta g_{1 x_j}\Big> \sim s^{-13}; \,\,\,\
\Big<\delta f_1 \delta f_{1 x_ix_j}\Big> \sim s^{-13};
\nonumber
\\
&&\Big<\delta f_1 \delta g_{1 x_ix_j}\Big> \sim s^{-13}; \,\,\,\
\Big<\delta g_1 \delta f_{1 x_ix_j}\Big> \sim s^{-13}; \,\,\,\
\Big<\delta g_1 \delta g_{1 x_ix_j}\Big> \sim s^{-13}; \,\,\,\
\Big<\delta f_{1 x_a} \delta f_{1 x_ix_j}\Big> \sim s^{-15};
\nonumber
\\
&&\Big<\delta f_{1 x_a} \delta g_{1 x_ix_j}\Big> \sim s^{-15}; \,\,\
\Big<\delta g_{1 x_a} \delta f_{1 x_ix_j}\Big> \sim s^{-15}; \,\,\
\Big<\delta g_{1 x_a} \delta g_{1 x_ix_j}\Big> \sim s^{-15}; \,\,\
\Big<\delta f_{1 x_a x_b} \delta f_{1 x_ix_j}\Big> \sim s^{-17};
\nonumber
\\
&&\Big<\delta f_{1 x_a x_b} \delta g_{1 x_ix_j}\Big> \sim s^{-17}; \,\,\
\Big<\delta g_{1 x_a x_b} \delta g_{1 x_ix_j}\Big> \sim s^{-17}~.
\end{eqnarray}
Using these, one can obtain the order of divergences as $T_2 \sim s^{-1}$, $T_3 \sim s^{-1}$ and so on. This implies that the second moment of the fluctuation of Ricci scalar diverges at least to the order of 
\begin{align}
\langle\d R_1\d R_1\rangle\sim s^{-1}~.
\end{align}
%%%%%%%%%%%%%%%%%%%%%%%%%%%%%%%%%%%%%%%%%%%%%%%%%%%%%%%%%%%%%%%%%%%%%%%%%%%%%%%%%%%

%--------------------------------------------------------------------------------
\subsection{Quevedo-II metric}
The Ricci scalar of Quevedo-II metric is given by \eqref{RICCI2}. The corresponding fluctuation in $R_2$ is,
\begin{eqnarray}
\delta R_2 &=& \frac{1}{2(f_2 g_2+h^2_2)^2}\Big[\delta f_2 \Big\{(f_{2Y} g_{2Y} - g^2_{2M}+2g_{2Y}h_{2M})-2g_2(f_{2YY}+h_{2MY}-g_{2MM})\Big\}
\nonumber 
\\
&& +\delta g_2\Big\{f^2_{2Y}+f_{2M}(2h_{2Y}-g_{2M})-2f_2(f_{2YY}+h_{2MY}-g_{2MM})\Big\}
\nonumber
\\
&& +\delta h_2 \Big\{-g_{2Y}f_{2M}+f_{2Y}(2h_{2Y}+g_{2M})+4h_{2Y}h_{2M}-2g_{2M}h_{2M}-2h_2(f_{2YY}+2h_{2MY}-g_{2MM})
\nonumber
\\
&& -2h_2(f_{2YY}+2h_{2MY}-g_{2MM})\Big\}+\delta f_{2M}\Big\{g_2(2h_{2Y}-g_{2M})-h_2 g_{2Y}\Big\} 
\nonumber
\\
&& +\delta g_{2M}\Big\{-2g_{2M}f_2 - f_{2M}g_2 +f_{2Y}h_2 - 2h_{2M}h_2\Big\}+ \delta h_{2M}\Big\{2g_{2Y}f_2+4h_{2Y}h_2-2g_{2M}h_2\Big\}
\nonumber
\\
&& +\delta f_{2Y}\Big\{f_2 g_{2Y}+2g_2f_{2Y}+h_2(2h_{2Y}+g_{2M})\Big\}+\delta g_{2Y}\Big\{f_{2Y}f_2+2h_{2M}f_2 -f_{2M}h_2\Big\} 
\nonumber
\\
&& +\delta h_{2Y}\Big\{2 f_{2M}g_2+2f_{2Y}h_2+4h_{2M}h_2 \Big\}+\delta g_{2MM}\Big\{-2f_2g_2+2h^2_2\Big\} \nonumber
\\
&& +\delta f_{2YY}\Big\{-2f_2g_2-2h^2_2\Big\}+\delta h_{2MY}\Big\{2f_2g_2-4h^2_2\Big\} 
\nonumber
\\
&& + \Big\{-\delta f_2\Big(\frac{g_2}{(f_2g_2+h^2_2)^3}\Big)-\delta g_2\Big(\frac{f_2}{(f_2g_2+h^2_2)^3}\Big)-\delta h_2\Big(\frac{h_2}{(f_2g_2+h^2_2)^3}\Big)\Big\}\Big[f_2(f_{2Y} g_{2Y}-g_{2M}^{2}+2g_{2Y}h_{2M})
\nonumber
\\
&&+g_2\Big\{f_{2Y}^2+f_{2M}(2h_{2Y}-g_{2M})-2f_2(f_{2YY}+h_{2MY}-g_{2MM})\Big\}
\nonumber
\\
&& +h_2\Big\{-g_{2Y}f_{2M}+f_{2Y}(2h_{2Y}+g_{2M})+4h_{2Y}h_{2M}-2g_{2M}h_{2M}-2h_2(f_{2YY}+2h_{2MY}-g_{2MM})\Big\}\Big]~, 
\label{delr2}
\end{eqnarray}
where $f_2=-[1+\b M S_{MM}+S_{MM}^2+S_{MY}^2]$, $g_2=1-\t X YS_{YY}+S_{YY}^2+S_{MY}^2$ and $h_2=\frac{1}{2}(\b M-\t X Y) S_{MY}+S_{MM}S_{MY}+S_{YM}S_{YY}$ as we have obtained earlier. Considering the variations we have,
\begin{eqnarray}
\delta f_2 &=& -\Big[M S_{MM}\delta \b +\b M \delta S_{MM}+ 2 S_{MM}\delta S_{MM}+2 S_{MY}\delta S_{MY}\Big]~;\\
\delta g_2 &=& \Big[-YS_{YY}\delta \t X - \t XY\delta S_{YY}+ 2S_{YY}\delta S_{YY}+ 2S_{MY}\delta S_{MY}\Big]~;\\
\delta h_2 &=& \frac{1}{2}\Big(M\delta \b- Y\delta \t X\Big)S_{MY}+\frac{1}{2}(\b M-\t X Y)\delta S_{MY} \nonumber
\\
&&+S_{MM}\delta S_{MY}+S_{MY}\delta S_{MM} +S_{MY}\delta S_{YY}+ S_{YY}\delta S_{MY}~;\\
\delta f_{2M} &=& -\Big[S_{MM}\delta \b + \b \delta S_{MM}+ M S_{MMM}\delta \b + \b M \delta S_{MMM}+ 2M S_{MM}\delta S_{MM} 
\nonumber
\\
&&+ 2S_{MM}\delta S_{MMM}+2S_{MMM}\delta S_{MM}+ 2S_{MY}\delta S_{MYM}+ 2S_{MYM}\delta S_{MY}\Big]~;
\end{eqnarray}
and so on for the variations in eqn \eqref{delr2}. Hence again calculating $\Big<\delta f_2 \delta f_2\Big>$, we see that the known divergence is from the quantitity $M^2S^2_{MM}\Big<(\delta \b)^2\Big>$ which is of the order $\sim s^{-9}.$ However we are unable at present to calculate the correlations of the other terms as per the presciption of the off-equlibrium linear stability analysis in \cite{Kaburakipla}. In the same vein, $\Big<\delta g_2 \delta g_2\big>$, $\Big<\delta h_2 \delta h_2\big>$ have a calculable divergence as $\sim s^{-9}.$ For the correlation with derivative terms we have, for example, $\Big<\delta f_2 \delta f_{2M}\big>$ which has a known divergence of $\sim s^{-11}$ and so on. It must be mentioned that terms like $\Big<\delta f_2 \delta g_2\Big>$ or the correlation of their deriavatives have a known/calculable divergence of zero since $\b$ and $\t X$ are independant parameters.

In order to compute the correlation in the fluctuations $\Big<\delta R_2 \delta R_2\Big>$ of the Ricci Scalar from the Quevedo metric(type $2$), we have from \eqref{delr2}, terms like,
\begin{equation}
\frac{1}{4(f_2g_2+h^2_2)^4}\Big<\delta f_2\delta f_2\Big>\Big\{(f_{2Y} g_{2Y} - g^2_{2M}+2g_{2Y}
h_{2M})-2g_2(f_{2YY}+
h_{2MY}-g_{2MM})\Big\}^2\\ \nonumber
\end{equation} 
which has a known/calculable order of $\mathcal{O}$($s^7$). Same analysis follows for the various self and cross terms in $\Big<\delta R_2 \delta R_2\Big>$ and it can be verified that they have either have a known/calculable order of $\mathcal{O}$($s^7$) or they vanish (due to the presence of cross terms like $\Big<\delta f_2 \delta g_2\Big>$).
Hence as such it cannot be said with certainty, whether the correlation of the fluctuations of the Ricci scalar ($\Big<\delta R_2 \delta R_2\Big>$) in the Quevedo metric type $2$ diverge or not. We have seen that the terms that can be calculated are indeed finite or they vanish. However the presence of terms like $\Big<\delta \b \delta S_{MM}\Big>$ and alike prevent us from making conclusions here about the divergence of the fluctuations.
%------------------------------------------------------------------------------------------------

\end{document}